\documentclass[12pt]{article}
\usepackage{geometry}                
\usepackage{graphicx}
\usepackage{bm}
\usepackage{amssymb,amsmath,amsthm}
\usepackage{mathrsfs} 
\usepackage{bm}
\usepackage{graphicx,epsfig}
\usepackage[dvips]{color}
\usepackage{amssymb,amsmath,amsthm,amssymb}

\def\J{{\mathscr{J}}}
\def\sech{{\mathrm{sech}}}

\def\TODAY{8 January 2009}

\title{\bf Analytic bounds on transmission probabilities}
\author{{\Large Petarpa Boonserm and Matt Visser}\\[5pt]
School of Mathematics, Statistics, and Operations Research\\
Victoria University of Wellington, New Zealand\\[5pt]
{\sf \small \{petarpa.boonserm,matt.visser\}@msor.vuw.ac.nz}  }
\date{\TODAY;  \LaTeX-ed  \today}                                           

\begin{document}
\maketitle
\def\d{{\mathrm{d}}}
\newcommand{\scri}{\mathscr{I}}
\newcommand{\sun}{\ensuremath{\odot}}
\numberwithin{equation}{section} 
\def\J{{\mathscr{J}}}
\def\sech{{\mathrm{sech}}}
\newtheorem{theorem}{Theorem}
\def\Re{{\mathrm{Re}}}
\def\Im{{\mathrm{Im}}}
\begin{abstract}

We develop some new analytic bounds on transmission probabilities (and the related reflection probabilities and Bogoliubov coefficients) for generic one-dimensional scattering problems.  To do so we rewrite the Schr\"odinger equation for some complicated potential whose properties we are trying to investigate in terms of some simpler potential whose properties are assumed known, plus a (possibly large) ``shift'' in the potential. Doing so permits us to extract considerable useful information without having to exactly solve the full scattering problem.

\vskip 10 pt

Pacs numbers: 03.65.-w, 03.65.Xp, 03.65.Nk

\vskip 10 pt




\end{abstract}

\maketitle
\newtheorem{lemma}{Lemma}
\newtheorem{corollary}{Corollary}
\newtheorem{definition}{Definition}
\def\d{{\mathrm{d}}}
\def\implies{\Rightarrow}
\def\arctanh{{\mathrm{arctanh}}}
\def\SIM{\triangleq}
\clearpage
\tableofcontents
\clearpage

\section{Introduction}

In several earlier papers~\cite{pra, ann-phys, greybody, jpa}, the present authors have derived a number of rigourous bounds on transmission probabilities (and reflection probabilities, and Bogoliubov coefficients) for one-dimensional scattering problems. The derivation of these bounds generally proceeds by rewriting the Schr\"odinger equation in terms of some equivalent system of first-order equations, and then analytically bounding the growth of certain quantities related to the net flux of particles as one sweeps across the potential. In the present article we shall obtain significantly different results, of both theoretical and practical interest.

While a vast amount of effort has gone into studying the Schr\"odinger equation and its scattering properties~\cite{dicke-wittke, merzbacher, landau, shankar, capri, messiah, branson-joachim, liboff, bohm, tunnelling, eberly, wkb-born, continued-fraction, geometric-transfer-matrix, inverse, aharonov, DE-4-T}, it appears that relatively little work has gone into providing general analytic bounds on the transmission probabilities, (as opposed to approximate estimates). The only known results as far as we have been able to determine are presented in~\cite{pra, ann-phys, greybody, jpa}. Several quite remarkable bounds were first derived in~\cite{pra}, with further discussion and an alternate proof being provided in~\cite{ann-phys}. These bounds were originally used by one of the present authors as a technical step when studying a specific model for sonoluminescence~\cite{sonoluminescence},  and since then have also been used to place limits on particle production in analogue spacetimes~\cite{analogue} and resonant cavities~\cite{cavity}, to investigate qubit master equations~\cite{qubit}, and to motivate further general investigations of one-dimensional scattering theory~\cite{1d-integral, 1d-vector, 1d-lippmann}. Recently, these bounds have also been applied to the greybody factors of a Schwarzschild black hole~\cite{greybody}. Most recently, significant extensions of the original bounds have been developed~\cite{jpa} by adapting the Miller--Good transformations~\cite{miller-good}.

In the current article we again return to this problem, developing a new set of techniques that are more amenable to the development of both \emph{upper} and \emph{lower} bounds on the transmission probabilities. For technical reasons the new techniques are also more amenable to investigating behavior ``under the barrier''.  The basic idea is to re-cast the Schr\"odinger equation for some complicated potential whose properties we are trying to investigate in terms of some simpler potential whose properties are assumed known, plus a (possibly large) ``shift'' in the potential.

\section{From Schr\"odinger equation to system of ODEs}

We are interested in the scattering properties of the Schr\"odinger equation,
\begin{equation}
\psi''(x) + k(x)^2 \; \psi(x) = 0,
\label{E:sde}
\end{equation}
where $k(x)^2 = 2m[E-V(x)]/\hbar^2$.  As long as $V(x)$ tends to finite (possibly distinct) constants $V_{\pm\infty}$ on left and right infinity, then for $E>\max\{V_{+\infty},V_{-\infty}\}$ one can set up a one-dimensional scattering problem in a completely standard manner --- see, for example, standard references such as~\cite{dicke-wittke, merzbacher, landau, shankar, capri, messiah, branson-joachim, liboff, bohm}.  The scattering problem is completely characterized by the transmission and reflection \emph{amplitudes} (denoted $t$ and $r$), although the most important aspects of the physics can be extracted from the transmission and reflection \emph{probabilities} ($T= |t|^2$ and $R=|r|^2$).

\subsection{Ansatz}

The idea is to try to say things about exact solutions to the ODE
\begin{equation}
\psi''(x) + k^2(x)\; \psi(x) = 0,
\end{equation}
by comparing this ODE to some ``simpler'' one
\begin{equation}
\psi_0''(x) + k_0^2(x)\; \psi_0(x) = 0,
\end{equation}
for which we are assumed to the know exact solutions $\psi_0(x)$.
In a manner similar to the analysis in references~\cite{pra,ann-phys}, we will start by introducing the ansatz
\begin{equation}
\label{E:representation}
\psi(x) = 
a(x) \;\psi_0(x)+ b(x) \;\psi_0^*(x).
\end{equation}
This representation is
of course extremely highly redundant, since \emph{one} complex number $\psi(x)$ has
been traded for \emph{two} complex numbers $a(x)$ and $b(x)$.  This redundancy allows us, without any loss of generality, to enforce one auxiliary constraint connecting $a(x)$ and $b(x)$. We find it particularly useful to enforce the auxiliary condition
\begin{equation}
\label{E:gauge}
{\d a\over \d x} \; \psi_0+ 
{\d b\over \d x} \; \psi_0^*
= 0.
\end{equation}
Subject to this auxiliary constraint on the derivatives of $a(x)$ and $b(x)$, the derivative of $\psi(x)$ takes on the especially simple form
\begin{equation}
\label{E:gradient}
{\d\psi\over \d x} = a \; \psi_0' + b \; \psi_0^*{}'.
\end{equation}
(This ansatz is largely inspired by the techniques of references~\cite{pra,ann-phys}, where JWKB estimates for the wave function were similarly used as a ``basis'' for formally writing down the exact solutions.)

\subsection{Probability density and probability current}

For the probability density we have:
\begin{eqnarray}
\rho  &=& \psi^* \psi
\\
&=& \big|  a(x) \psi_0
+  b(x) \psi_0^* \big|^2
\\
&=& \{ |a|^2 + |b|^2| \} |\psi_0|^2 + 2 \mathrm{Re} \, \{a b^* \psi_0^2  \} 
\\
&=& \{ |a|^2 + |b|^2| \} \rho_0 + 2 \mathrm{Re} \, \{a b^* \psi_0^2  \}.
\end{eqnarray}
Furthermore, for the probability current:
\begin{eqnarray}
\mathscr{J} &=& \mathrm{Im} \, \bigg\{ \psi^* {\d \psi \over \d x}\bigg\}
\\
&=& \mathrm{Im} \, \bigg\{ \left[ a^* \psi_0^* + b^* \psi_0 \right] \; 
\left[a \psi_0' + b \psi_0^*{}'  \right] \bigg\}\qquad\;
\\
&=&\mathrm{Im} \, \Bigg\{ |a|^2 \psi_0^* \psi_0' + |b|^2 \psi_0 \psi_0^*{}' + ab^* \psi_0 \psi_0' + a^*b \psi_0^*\psi_0^*{}'
 \Bigg\} 
 \\
&=& \{ |a|^2- |b|^2 \} \; \mathrm{Im} \, \{ \psi_0^* \; \psi_0'\}
\\
&=&   \{ |a|^2- |b|^2 \}  \; \mathscr{J}_0.
\end{eqnarray}
Under the conditions we are interested in, (corresponding to a time-independent solution of the Schr\"odinger equation), we have $\dot\rho=0$, and so $\partial_x \mathscr{J}=0$. (And similarly  $\dot\rho_0=0$, so $\partial_x \mathscr{J}_0=0$.) That is, $\mathscr{J}$ and   $\mathscr{J}_0$ are position-independent constants,  which then puts a constraint on the amplitudes $|a|$ and $|b|$.  Applying an appropriate boundary condition, which we can take to be $a(-\infty)=1$, $b(-\infty)=0$, we then see
\begin{equation}
 |a|^2 - |b|^2 = 1.
\end{equation}
This observation justifies interpreting $a(x)$ and $b(x)$ as ``position-dependent Bogoliubov coefficients''. Furthermore without any loss in generality we can choose the normalizations on $\psi$ and $\psi_0$ so as to set the net fluxes to unity: $\mathscr{J}=\mathscr{J}_0=1$.

\subsection{Second derivatives of the wavefunction}
%
We shall now re-write the Schr\"odinger equation in terms of two coupled
first-order differential equations for these position-dependent
Bogoliubov coefficients $a(x)$ and $b(x)$. To do this, evaluate $\d^2\psi/ \d x^2$ making
repeated use of the auxiliary condition
\begin{eqnarray}
\label{E:double-gradient}
{\d^2\psi\over \d x^2} 
&=& 
{\d\over \d x} 
\left( a \,\psi_0' + b \, \psi_0^*{}' \right)
\\
&=&
a'  \, \psi_0' + b'  \, \psi_0^*{}' +  a  \, \psi_0'' + b \, \psi_0^*{}''
\\
&=& 
a'  \, \psi_0' - a'  \,  {\psi_0\over\psi_0^*}  \,  \psi_0^*{}' -  a  \, k_0^2  \, \psi_0 - b  \, k_0^2  \,  \psi_0^*
\\
&=& {a'\over\psi_0^*}  \, \{ \psi_0^* \psi_0' - \psi_0 \psi_0^*{}' \} 
-  
k_0^2  \, [ a  \psi_0 + b \psi_0^*] 
\\
&=& {2i\J_0 a'\over \psi_0^*} 
-  
k_0^2  \,  [ a  \psi_0 + b \psi_0^*] 
\\
&=& {2i a'\over \psi_0^*} 
-  
k_0^2  \,  [ a  \psi_0 + b \psi_0^*].
\end{eqnarray}
Where in the last line we have finally used our normalization choice $\J_0=1$.
This is one of the two relations we wish to establish. 
Now use the gauge condition to eliminate $\d a/\d x$ in favour of $\d b/\d x$ to obtain a second relation for  $\d^2\psi/ \d x^2$.  This now permits us to write $\d^2\psi/ \d x^2$ in
either of the two equivalent forms
\begin{eqnarray}
\label{E:double-gradient2}
{\d^2\psi\over \d x^2} 
&=& {2i a'\over \psi_0^*} 
-  
k_0^2 \,  [ a  \psi_0 + b \psi_0^*];
\\
&=& - {2i b'\over \psi_0} 
-  
k_0^2 \,  [ a  \psi_0 + b \psi_0^*].
\end{eqnarray}
%

\subsection{SDE as a first-order system}

Now insert these formulae for the second derivative of the wavefunction into the Schr\"odinger equation written in the
form
\begin{equation}
{\d^2\psi\over \d x^2} + k(x)^2 \; \psi  = 0,
\end{equation}
to deduce the \emph{pair} of first-order ODEs:
\begin{eqnarray}
\label{E:system-a}
{\d a\over \d x} &=& +
{i\over2} [k^2-k_0^2]\;
\{ a\; |\psi_0|^2 + b\; \psi_0^*{}^2 
\};
\\
\label{E:system-b}
{\d b\over \d x} &=& -
{i\over2} [k^2-k_0^2]\;
\{ a \;\psi_0^2 + b \; |\psi_0|^2 \}.
\end{eqnarray}
It is easy to verify that this first-order system is compatible with
the auxiliary condition (\ref{E:gauge}), and that by iterating the
system twice (subject to this auxiliary condition) one recovers exactly
the original Schr\"odinger equation. 
We can re-write this 1st-order system of ODEs in matrix form as
\begin{equation}
{\d\over \d x} \left[\begin{matrix} a \\ b\end{matrix}\right] = 
{i[k^2-k_0^2]\over2} 
\left[
\begin{matrix}
|\psi_0|^2 & \psi_0^*{}^2\\
-\psi_0^2 & - |\psi_0|^2
\end{matrix}
\right]
\left[ \begin{matrix} a \cr b\end{matrix}\right].
\end{equation}
(Matrix ODEs of this general form are often referred to as Shabhat--Zakharov  or Zakharov--Shabat systems~\cite{pra}. This matrix ODE can be used to write down a formal solution to the SDE in terms of ``path-ordered exponentials'' as in references~\cite{pra,ann-phys}.  We choose not to adopt this route here, instead opting for a more direct computation in terms of the magnitudes and phases of $a$ and $b$.)

\subsection{Formal (partial) solution}

Define magnitudes and  phases by
\begin{equation}
a = |a|\; e^{i\phi_a}; \qquad b = |b|\; e^{i\phi_b}; \qquad \psi_0 = |\psi_0|\; e^{i\phi_0}.
\end{equation}
Calculate
\begin{equation}
a' =  |a|' \, e^{i\phi_a} +  i |a| \, e^{i\phi_a} \, \phi_a' = e^{i\phi_a} \left\{ |a|' + i |a| \, \phi_a'\right\},
\end{equation}
whence
\begin{equation}
|a|' + i |a| \, \phi_a' = {i\over2} [k^2-k_0^2]\;  |\psi_0|^2 \; \{ |a|\; + |b|\; e^{-i(\phi_a-\phi_b+2\phi_0)} \}.
\label{E:a}
\end{equation}
Similarly we also have
\begin{equation}
|b|' + i |b| \, \phi_b' = -{i\over2} [k^2-k_0^2]\;  |\psi_0|^2 \; \{ |b|\; + |a|\; e^{-i(\phi_b-\phi_a-2\phi_0)} \}.
\label{E:b}
\end{equation}
Now take the real part of both these equations, 
whence
\begin{equation}
|a|' = +{1\over2} [k^2-k_0^2] \;  |b| \; |\psi_0|^2  \sin(\phi_a-\phi_b+2\phi_0);
\end{equation}
\begin{equation}
|b|' = +{1\over2} [k^2-k_0^2] \;  |a| \; |\psi_0|^2  \sin(\phi_a-\phi_b+2\phi_0).
\end{equation}
Therefore
\begin{equation}
|a|' =   {1\over2} [k^2-k_0^2] \; |\psi_0|^2  \sin(\phi_a-\phi_b+2\phi_0)  \;   \sqrt{|a|^2-1}.
\end{equation}
That is
\begin{equation}
{|a|' \over \sqrt{|a|^2-1}}  =  {1\over2} [k^2-k_0^2] \; |\psi_0|^2  \sin(\phi_a-\phi_b+2\phi_0),
\end{equation}
whence
\begin{equation}
\left\{ \cosh^{-1} |a| \right\}_{x_1}^{x_2}  = {1\over2}
 \int _{x_1}^{x_2} [k^2-k_0^2] \; |\psi_0|^2  \sin(\phi_a-\phi_b+2\phi_0) \; \d x.
\end{equation}
Now apply the boundary conditions: At $x=-\infty$ we have both $a(-\infty)=1$, and $b(-\infty)=0$. Therefore
\begin{equation}
\cosh^{-1} |a(x)|  = {1\over2} 
 \int _{-\infty}^{x}  [k^2-k_0^2] \; |\psi_0|^2  \sin(\phi_a-\phi_b+2\phi_0) \; \d x,
\end{equation}
and so
\begin{equation}
\label{E:pre-Theta}
|a(x)|  = \cosh\left\{ {1\over2} 
 \int _{-\infty}^{x} [k^2-k_0^2] \; |\psi_0|^2  \sin(\phi_a-\phi_b+2\phi_0) \; \d x \right\}.
\end{equation}
In particular
\begin{equation}
\cosh^{-1} |a(\infty)|  = {1\over2} 
 \int _{-\infty}^{+\infty}  [k^2-k_0^2] \; |\psi_0|^2  \sin(\phi_a-\phi_b+2\phi_0) \; \d x,
\end{equation}
or equivalently
\begin{equation}
|a(\infty)|  = \cosh\left\{ {1\over2} 
 \int _{-\infty}^{+\infty} [k^2-k_0^2] \; |\psi_0|^2  \sin(\phi_a-\phi_b+2\phi_0) \; \d x \right\}.
\end{equation}
Of course this is only a \emph{formal} solution since $\phi_a(x)$ and $\phi_b(x)$ are, (at least at this stage), ``unknown''. But we shall argue that this formula still contains useful information. In particular, in view of the normalization conditions relating $a$ and $b$, and the parity properties of $\cosh$ and $\sinh$,  we can also write
\begin{equation}
\label{E:a1}
|a(\infty)|  = \cosh\left| {1\over2} 
 \int _{-\infty}^{+\infty} [k^2-k_0^2] \; |\psi_0|^2  \sin(\phi_a-\phi_b+2\phi_0) \; \d x \right|;
\end{equation}
\begin{equation}
\label{E:b1}
|b(\infty)|  = \sinh\left| {1\over2} 
 \int _{-\infty}^{+\infty} [k^2-k_0^2] \; |\psi_0|^2  \sin(\phi_a-\phi_b+2\phi_0) \; \d x \right|.
\end{equation}

\subsection{First set of bounds}

To determine the first elementary set of bounds on $a$ and $b$ is now trivial. We just note that
\begin{equation}
|\sin(\phi_a-\phi_b+2\phi_0)| \leq 1.
\end{equation}
Therefore
\begin{equation}
|a(\infty)|  \leq \cosh\left\{ {1\over2} 
 \int _{-\infty}^{+\infty} {|k^2-k_0^2| \;|\psi_0|^2} \, \d x \right\};
\end{equation}
\begin{equation}
|b(\infty)|  \leq \sinh\left\{ {1\over2} 
 \int _{-\infty}^{+\infty} {|k^2-k_0^2| \;|\psi_0|^2} \, \d x \right\}.
\end{equation}
What does this now tell us about the Bogoliubov coefficients?

\subsection{Bogoliubov coefficients}

The slightly unusual thing, (compared to our earlier work in references~\cite{pra,ann-phys,jpa}),  is that now the ``known'' function $\psi_0$ may also have its own Bogoliubov coefficients. Let us assume we have set our boundary conditions so that for the ``known'' situation 
\begin{equation}
\psi_0(x\approx - \infty) \sim \exp\{ik(-\infty) x\},
\end{equation}
and
\begin{equation}
\psi_0(x\approx + \infty) \sim \alpha_0 \exp\{ik(+\infty) x\} 
+ \beta_0  \exp\{-ik(+\infty) x\}.
\end{equation}
Then the way we have set things up, for the ``full'' problem we still have
\begin{equation}
\psi(x\approx - \infty) \sim \exp\{ik(-\infty) x\},
\end{equation}
whereas
\begin{eqnarray}
\psi(x\approx + \infty) &\sim& a(\infty) \, \psi_0(x) + b(\infty) \, \psi_0^*(x) \\
&\sim& 
[\alpha_0 \,  a(\infty) + \beta_0^* \, b(\infty) ] \,\exp\{ik(+\infty) x\} 
\nonumber\\
&&
\qquad 
+ [\beta_0 \, a(\infty) + \alpha_0^* \, b(\infty) ] \, \exp\{-ik(+\infty) x\}.
\end{eqnarray}
That is, the overall Bogoliubov coefficients satisfy
\begin{equation}
\alpha = \alpha_0 \, a(\infty) + \beta_0^* \, b(\infty);
\end{equation}
\begin{equation}
\beta = \beta_0 \, a(\infty) + \alpha_0^* \, b(\infty).
\end{equation}
These equations relate the Bogoliubov coefficients of the ``full'' problem $\{\psi(x),\,k(x)\}$ to those of the simpler ``known'' problem $\{\psi_0(x),\,k_0(x)\}$, plus the evolution of the $a(x)$ and $b(x)$ coefficients.
Now observe that
\begin{equation}
|\alpha| \leq |\alpha_0| \; |a(\infty)| + |\beta_0| \; |b(\infty)|.
\end{equation}
But we can define
\begin{equation}
|\alpha_0| = \cosh \Theta_0; \quad |\beta_0| = \sinh \Theta_0; 
\qquad 
 |a(\infty)| = \cosh \Theta; \quad  |b(\infty)| = \sinh \Theta; 
\end{equation}
in terms of which 
\begin{equation}
|\alpha| \leq \cosh \Theta_0 \cosh \Theta +\sinh \Theta_0 \sinh \Theta
= \cosh\left(\Theta_0 + \Theta\right).
\end{equation}
That is:  Since we know 
\begin{equation}
\Theta \leq \Theta_\mathrm{bound} 
\equiv {1\over2}   \int _{-\infty}^{+\infty} {|k^2-k_0^2| \;|\psi_0|^2} \d x,
\end{equation}
we can deduce
\begin{equation}
|\alpha|  \leq \cosh\left\{ \cosh^{-1}|\alpha_0| + {1\over2} 
 \int _{-\infty}^{+\infty} {|k^2-k_0^2| \;|\psi_0|^2} \d x \right\};
\end{equation}
\begin{equation}
|\beta|  \leq \sinh\left\{ \sinh^{-1}|\beta_0| + {1\over2} 
 \int _{-\infty}^{+\infty} {|k^2-k_0^2| \;|\psi_0|^2} \d x \right\}.
\end{equation}
%

\subsection{Second set of bounds}

A considerably trickier inequality, now leading to a \emph{lower bound} on the Bogoliubov coefficients, is obtained by considering what the phases would have to be to achieve as much destructive interference as possible. That implies \begin{equation}
|\alpha| \geq |\alpha_0| \; |a(\infty)| - |\beta_0| \; |b(\infty)|,
\end{equation}
whence
\begin{equation}
|\alpha| \geq 
\cosh\left|\Theta_0 - \Theta\right|.
\end{equation}
Therefore, using $\Theta \leq \Theta_\mathrm{bound}$, it follows that \emph{as long as} $ \Theta_\mathrm{bound} < \Theta_0$, one can deduce 
\begin{equation}
|\alpha| \geq 
\cosh\left\{\Theta_0 - \Theta_\mathrm{bound}\right\}.
\end{equation}
(If on the other hand $\Theta_\mathrm{bound} \geq \Theta_0$, then one only obtains the trivial bound $|\alpha|\geq 1$.) Another way of writing these bounds is as follows
\begin{equation}
|\alpha|  \geq \cosh\left\{ \cosh^{-1}|\alpha_0| - {1\over2} 
 \int _{-\infty}^{+\infty} {|k^2-k_0^2| \;|\psi_0|^2} \d x \right\};
\end{equation}
\begin{equation}
|\beta|  \geq \sinh\left\{ \sinh^{-1}|\beta_0| - {1\over2} 
 \int _{-\infty}^{+\infty} {|k^2-k_0^2| \;|\psi_0|^2} \d x \right\};
\end{equation}
with the tacit understanding that the bound remains valid only so long as argument of the hyperbolic function is positive.

\subsection{Transmission probabilities}

As usual, the transmission probability (barrier penetration probability) is related to the Bogoliubov coefficient by
\begin{equation}
T = {1\over|\alpha|^2},
\end{equation}
whence
\begin{equation}
T  \geq \sech^2\left\{ \cosh^{-1}|\alpha_0| + {1\over2} 
 \int _{-\infty}^{+\infty} {|k^2-k_0^2| \;|\psi_0|^2} \d x \right\}.
\end{equation}
That is
\begin{equation}
T  \geq \sech^2\left\{ \cosh^{-1}(T_0^{-1/2}) + {1\over2} 
 \int _{-\infty}^{+\infty} {|k^2-k_0^2| \;|\psi_0|^2} \d x \right\},
\end{equation}
or even
\begin{equation}
T  \geq \sech^2\left\{ \sech^{-1}(T_0^{1/2}) + {1\over2} 
 \int _{-\infty}^{+\infty} {|k^2-k_0^2| \;|\psi_0|^2} \d x \right\}.
\end{equation}
Furthermore, as long as the argument of the $\sech$ is positive, we also have the upper bound
\begin{equation}
T  \leq \sech^2\left\{ \sech^{-1}(T_0^{1/2}) - {1\over2} 
 \int _{-\infty}^{+\infty} {|k^2-k_0^2| \;|\psi_0|^2} \d x \right\}.
\end{equation}
If one wishes to make the algebraic dependence on $T_0$ clearer, by expanding the hyperbolic functions these formulae may be recast as
\begin{equation}
T  \geq { T_0 \over 
\left[ \cosh\left\{ {1\over2} \int _{-\infty}^{+\infty} {|k^2-k_0^2| \;|\psi_0|^2} \d x \right\} 
+ \sqrt{1-T_0}  \sinh\left\{ {1\over2} \int _{-\infty}^{+\infty} {|k^2-k_0^2| \;|\psi_0|^2} \d x \right\} \right]^2},
\end{equation}
and (as long as the numerator is positive before squaring) 
\begin{equation}
T  \leq { T_0 \over 
\left[ \cosh\left\{ {1\over2} \int _{-\infty}^{+\infty} {|k^2-k_0^2| \;|\psi_0|^2} \d x \right\} 
- \sqrt{1-T_0}  \sinh\left\{ {1\over2} \int _{-\infty}^{+\infty} {|k^2-k_0^2| \;|\psi_0|^2} \d x \right\} \right]^2}.
\end{equation}

\section{Consistency check}

There is one special case in which we can easily compare with the previous results of references~\cite{pra,ann-phys}. Take $k_0 = k(\pm\infty)$ to be independent of position, so that our comparison problem is a free particle. In that case 
\begin{equation}
\psi_0 = {\exp(ik_0 x)\over\sqrt{k_0}}; \qquad
|\psi_0|^2 = {1\over k_0} ; \qquad \J_0 = 1; \quad \alpha_0 = 1; \qquad \beta_0 = 0.
\end{equation}
Then the bounds derived above simplify to
\begin{equation}
|\alpha|  \leq \cosh\left\{  {1\over2k_0} 
 \int _{-\infty}^{+\infty} |k^2-k_0^2| \;\d x \right\},
\end{equation}
\begin{equation}
|\beta|  \leq \sinh\left\{ {1\over2k_0} 
 \int _{-\infty}^{+\infty} |k^2-k_0^2| \;\d x \right\}.
\end{equation}
This is  ``Case I'' of reference~\cite{pra} and the ``elementary bound'' of reference~\cite{ann-phys}, which demonstrates consistency whenever the formalisms overlap. 
(Note that it is not possible to obtain ``Case II'' of reference~\cite{pra} or the ``general bound'' of reference~\cite{pra, ann-phys}  from the present analysis --- this is not a problem, it is just an indication that this new bound really is a \emph{different} bound that only partially overlaps with the previous results of references~\cite{pra, ann-phys, jpa}.

A second (elementary) check is to see what happens if we set $\psi(x)\to\psi_0(x)$, effectively assuming that the full problem is analytically solvable. In that case $T\to T_0$,  (and similarly both $\alpha\to\alpha_0$ and $\beta\to\beta_0$), as indeed they should.

\section{Keeping the phases?}
We can extract a little more information by taking the imaginary parts of equations (\ref{E:a}) and (\ref{E:b}) to obtain:
\begin{equation}
\phi_a' = {1\over2} [k^2-k_0^2]\;  |\psi_0|^2 \; \left\{ 1\; + {|b|\over |a|} \; \cos(\phi_a-\phi_b+2\phi_0) \right\};
\label{E:aa}
\end{equation}
\begin{equation}
\phi_b' = -{1\over2} [k^2-k_0^2]\;  |\psi_0|^2 \; \left\{ 1 + {|a|\over |b|} \; \cos(\phi_b-\phi_a-2\phi_0) \right\}.
\label{E:bb}
\end{equation}
Subtracting
\begin{equation}
(\phi_a-\phi_b)' = [k^2-k_0^2]\;  |\psi_0|^2 \; \left\{ 1\; + {1\over 2 |a| \, |b|} \; \cos(\phi_a-\phi_b+2\phi_0) \right\}.
\label{E:ab}
\end{equation}
This is now a differential equation that only depends on the difference in the phases --- the overall average phase $(\phi_a+\phi_b)/2$ has completely decoupled. (Likewise, in determining the transmission and reflection probabilities, this average phase also neatly decouples). 
To see how far we can push this observation, let us now define a ``nett'' phase
\begin{equation}
\Delta  = \phi_a-\phi_b+2\phi_0.
\end{equation}
Furthermore,  as per the previous subsections, we retain the definitions
\begin{equation}
|a| = \cosh\Theta; \qquad |b| = \sinh\Theta.
\end{equation}
Then equation (\ref{E:pre-Theta}) becomes
\begin{equation}
\label{E:Theta}
\Theta(x)  = \left\{ {1\over2} 
 \int _{-\infty}^{x} [k^2-k_0^2] \; |\psi_0|^2  \sin(\Delta(x)) \; \d x \right\}.
\end{equation}
while the ``nett'' phase satisfies
\begin{equation}
\Delta(x)' = \left\{ [k^2-k_0^2]\;  |\psi_0|^2 \; + 2\phi_0' \right\} + { [k^2-k_0^2]\;  |\psi_0|^2 \over \sinh[2\Theta(x)]} \; \cos[\Delta(x)] .
\label{E:ab2}
\end{equation}
We can even substitute for $\Theta(x)$ and thus rewrite this as a single \emph{integro-differential} equation for $\Delta(x)$:
\begin{equation}
\Delta(x)' = \left\{ [k^2-k_0^2]\;  |\psi_0|^2 \; + 2\phi_0' \right\} + 
{ [k^2-k_0^2]\;  |\psi_0|^2 \over 
\sinh\left(  \int _{-\infty}^{x} [k^2-k_0^2] \; |\psi_0|^2  \sin[\Delta(x)] \; \d x  \right)} \; \cos[\Delta(x)].
\label{E:ab3}
\end{equation}
This equation is completely equivalent to the original Schr\"odinger equation we started from. Unfortunately further manipulations seem intractable, and it does not appear practicable to push these observations any further.

\section{Application: Small shift in the potential}

Let us now consider the situation
\begin{equation}
V(x) = V_0(x) + \epsilon\; \delta V(x),
\end{equation}
for $\epsilon$ ``sufficiently small''.  

\subsection{First-order changes}
To be consistent with previous notation let us define 
\begin{equation}
k^2 = k_0^2 + \epsilon \; \left\{ {2m \; \delta V\over\hbar}\right\} \equiv k_0^2 + \epsilon\; \delta v.
\end{equation}
Using equation (\ref{E:pre-Theta}) we obtain the preliminary estimates
\begin{equation}
|a(x)|  = 1+ \mathcal{O}(\epsilon^2),
\end{equation}
and similarly 
\begin{equation}
|b(x)|  = \mathcal{O}(\epsilon).
\end{equation}
It is now useful to change variables by introducing some explicit phases so as to define
\begin{equation}
a = \tilde a \; \exp\left(+{i\over2} \int [k^2-k_0^2]\, |\psi_0^2|\, \d x\right); 
\end{equation}
\begin{equation}
b = \tilde b \; \exp\left(-{i\over2} \int [k^2-k_0^2]\, |\psi_0^2|\, \d x\right).
\end{equation}
Doing so modifies the system of differential equations (\ref{E:system-a}, \ref{E:system-b}) so that it becomes
\begin{eqnarray}
\label{E:system-aa}
{\d \tilde a\over \d x} &=& 
+{i\over2} [k^2-k_0^2] \; \tilde b\; \psi_0^*{}^2 \; \exp\left(-i \int [k^2-k_0^2] \, |\psi_0^2|\, \d x\right);
\\
\label{E:system-bb}
{\d \tilde b\over \d x} &=& 
- {i\over2} [k^2-k_0^2]\; \tilde a \;\psi_0^2 \; \exp\left(+i \int [k^2-k_0^2] \, |\psi_0^2|\, \d x\right).
\end{eqnarray}
The advantage of doing this is that in the current situation we can now estimate
\begin{eqnarray}
\label{E:system-aaa }
{\d \tilde a\over \d x} &=&  \mathcal{O}(\epsilon^2),
\\
\label{E:system-bbb}
{\d \tilde b\over \d x} &=& 
-  {i\epsilon\over2} \; \delta v(x) \;\psi_0^2(x) \; \exp\left(+i \int \epsilon \; \delta v\; |\psi_0^2| \; \d x\right) +  \mathcal{O}(\epsilon^3).
\end{eqnarray}
Integrating
\begin{equation}
\label{E:b1b}
\tilde b(\infty)  =  -  {i\epsilon\over2} 
 \int _{-\infty}^{+\infty} \delta v(x) \; \psi_0^2(x)  \; \exp\left(+i \int \epsilon \; \delta v\; |\psi_0^2| \; \d x\right)  \; \d x  + \mathcal{O}(\epsilon^3).
\end{equation}
This is not the standard Born approximation, though it can be viewed as an instance of the so-called ``distorted Born wave approximation''.  In terms of the absolute values we definitely have
\begin{equation}
|\tilde b(\infty)| =  |b(\infty)| \leq  \; {\epsilon\over2} \;
 \int _{-\infty}^{+\infty} |\delta v(x)| \; |\psi_0^2(x)|  \; \d x  + \mathcal{O}(\epsilon^3).
\end{equation}

\subsection{Particle production}

When it comes to considering particle production we note that
\begin{equation}
\beta =\beta_0 \; a(\infty)+ \alpha_0^* \; b(\infty)  = \beta_0 + \alpha_0^* \; b(\infty)  + \mathcal{O}(\epsilon^2),
\end{equation}
so the \emph{change} in the number of particles produced is
\begin{equation}
\delta|\beta^2| = \Re\left\{ 2 \alpha_0^* \, \beta_0\; b(\infty) \right\} +  \mathcal{O}(\epsilon^2).
\end{equation}
In particular
\begin{equation}
\left| \,\delta N \,\right| \leq  \epsilon \, |\alpha_0| \, |\beta_0|\;  \int _{-\infty}^{+\infty} |\delta v(x)| \; |\psi_0^2(x)| \;  \d x  +  \mathcal{O}(\epsilon^2),
\end{equation}
which we can also write as
\begin{equation}
\left| \,\delta N \,\right|\leq  \epsilon \;\sqrt{N_0(N_0+1)} \;  \int _{-\infty}^{+\infty} |\delta v(x)| \; |\psi_0^2(x)| \;  \d x  +  \mathcal{O}(\epsilon^2).
\end{equation}
Note that one will only get an order $\epsilon$ change in the particle production if the ``known'' problem $\{\psi_0,\, k_0\}$ already results in nonzero particle production.  

\subsection{Transmission probability}

To see how a small shift in the potential affects the transmission probability we note
\begin{equation}
T = {1\over|\alpha|^2} = {1\over| \alpha_0 \, a(\infty) + \beta_0^* \, b(\infty)|^2} =  {1\over| \alpha_0  + \beta_0^* \, b(\infty) + \mathcal{O}(\epsilon^2)|^2}.
\end{equation}
But then
\begin{equation}
T =  {1\over| \alpha_0 |^2 \; \left|1 + \{\beta_0^* \, b(\infty)/\alpha_0\} + \mathcal{O}(\epsilon^2)\right|^2},
\end{equation}
implying
\begin{equation}
T =   T_0 \;  \left\{1  - 2 \Re\left\{{\beta_0^* \, b(\infty)\over\alpha_0}\right\} + \mathcal{O}(\epsilon^2) \right\}.
\end{equation}
So the \emph{change} in the transmission probability is
\begin{equation}
\delta T =   - T_0 \;  \left\{2 \Re\left\{{\beta_0^* \; b(\infty)\over\alpha_0}\right\} + \mathcal{O}(\epsilon^2) \right\}.
\end{equation}
Taking absolute values one obtains
\begin{equation}
|\delta T| \leq  \epsilon \; T_0 \; \sqrt{1-T_0} \, \int _{-\infty}^{+\infty} |\delta v(x)| \; |\psi_0^2(x)| \;  \d x+ \mathcal{O}(\epsilon^2).
\end{equation}
Note that one will only get an order $\epsilon$ change in the transmission probability if the ``known'' problem $\{\psi_0,\, k_0\}$ already results in nonzero transmission (and nonzero reflection).

\section{Discussion}

What are the advantages of the particular bounds derived in this article?
\begin{itemize}

\item They are very simple to derive --- the algebra is a lot less complicated than some of the other approaches that have been developed~\cite{pra,ann-phys,greybody,jpa}. (And a lot less complicated than some of the blind alleys we have explored.)

\item Under suitable circumstances the procedure of this article yields both upper and lower bounds. Obtaining both upper and lower bounds is in general very difficult to do --- see in particular the attempts in~\cite{ann-phys}.

\item All of the other bounds we have developed~\cite{pra,ann-phys,greybody,jpa}  needed some condition on the phase of the wave-function, (some condition similar to $\varphi'\neq 0$), which had the ultimate effect of making it difficult to make statements about tunnelling ``under the barrier''.  There is no such requirement in the present analysis. (The closest analogue is that we need $\J_0\neq0$, which we normalize without loss of generality to $\J_0=1$.) In particular this means that there should be no particular difficulty in applying the bound in the classically forbidden region --- the ``art'' will lie in finding a suitable form for $\psi_0$ which is simple enough to carry out exact computations while still providing useful information.

\end{itemize}
In closing, we reiterate the fact that generic one-dimensional scattering problems, which have been extensively studied for close to a century,  nevertheless still lead to interesting features and novel results.

\section*{Acknowledgments}
This research was supported by the Marsden Fund administered by the Royal Society of New Zealand. PB was additionally supported by a scholarship from the Royal Government of Thailand. 


\end{document}